# MEASURING EXPECTATIONS IN OPTIONS MARKETS: AN APPLICATION TO THE S&P500 INDEX

ABEL RODRÍGUEZ AND ENRIQUE TER HORST

ABSTRACT. Extracting market expectations has always been an important issue when making national policies and investment decisions in financial markets. In option markets, the most popular way has been to extract implied volatilities to assess the future variability of the underlying with the use of the Black & Scholes formula. In this manuscript, we propose a novel way to extract the whole time varying distribution of the market implied asset price from option prices. We use a Bayesian nonparametric method that makes use of the Sethuraman representation for Dirichlet processes to take into account the evolution of distributions in time. As an illustration, we present the analysis of options on the S&P500 index.

## 1. INTRODUCTION

Derivatives have a significant influence on the behavior of spot markets. Understanding the expectations of actors in derivative markets can provide helpful insights on general economic conditions and the future behavior of the corresponding spot prices (French, 1986; Fama and French, 1987; Tomek, 1997). In futures markets, it is common to use the observed prices of the contracts together with a no-arbitrage condition to estimate the implied prices of the underlying asset in the spot market. In this case, the implied price corresponds to the net present value of a future contract. This implied

*Key words and phrases.* Nonparametric Bayes; Dependent Dirichlet process; European Options, Implied Prices.

We wish to thank Lars Stentoft and the Centre for Analytical Finance, University of Aarhus, Denmark, for providing us this data, as well as Samuel Malone for insightful conversations.

price, which is the fair price at inception of the derivative contract, can often be different from the price actually observed in the spot market.

Another useful example is structural credit risk. Extracting the unobserved implied asset price is a useful and necessary task in this field, whose purpose is assessing a corporation's credit risk. The central distinguishing point of structural credit models is the view of debt, equity, and other claims issued by a firm as option derivatives on the firms asset value (Black and Scholes, 1973; Merton, 1973, 1974). Given that the asset is unobserved, one has to estimate it in order to compute many important financial ratios that assess the financial risk of a corporation such as the probability of default, the corporation's bond spread, as well as its distance-to-default. The distance to default, defined as the distance (measured in standard deviations) of a firms asset value from its default threshold, has become popular as a measure of a firm's credit worthiness (Vassalou and Xing, 2004). A popular implementation of Merton's structural model Merton (1974) is the commercial KMV model, which through the use of the Black and Scholes pricing formula, is able to solve for the value of the asset given the equity of the firm, which is a call option. However, it depends critically on the assumptions underlying the Black and Scholes model, in particular, that prices follow a geometric Brownian motion.

Outside of the previous context, extracting the underlying/asset value is rarely done in financial option markets; instead, the implied volatility of returns from the Black and Scholes model is typically computed. Although this quantity is useful for understanding option prices, it also depends crucially on the lognormality assumption of returns and provides no real information about under/over valuation of assets in the spot market.

In principle, no-arbitrage conditions together with interest rates, call and put prices can be used to determine option-implied prices for the underlying asset in a similar fashion as for futures and forwards. However, one of the difficulties of using these prices is that, unlike in future markets, multiple observations are available for any given expiration date (corresponding to a different strike price), each leading to a slightly different implied price. This means that we need to deal with a *distribution* of implied prices, which can be highly non-Gaussian, presenting heavy tails and multimodality. The lack of normality implies that simple models based on summaries of the distribution (like the first two or three empirical moments) are inappropriate and can be misleading. Another difficulty is that a variable number of transactions takes place each day, as some strikes might not be traded. Therefore, independent density estimates for each day can be highly unstable, especially in periods of low liquidity when few transactions take place.

In this paper, we propose a Bayesian dynamic nonparametric model to estimate the collection of distribution of option-implied prices. This has numerous advantages over models based on empirical moments, as they are able to explicitly capture features like multimodality and allow us to estimate probabilities of default events that are unavailable under simpler models. Our model is based on the Dirichlet process (Ferguson, 1973, 1974), and is an extension of the dynamic Dirichlet process models in Rodriguez and ter Horst (2008) that incorporates stochastic volatility. It uses an infinite mixture of time evolving distributions, leading to a flexible and efficient model that borrows information across time periods to improve estimation and prediction. As a motivating example, we analyze European option and spot data for the S&P500 between January 1993 and March 1994. However, our argument can be immediately extended to other markets and other types of options.

Some possible alternatives to our model that have been extensively discussed in the literature include GARCH and stochastic volatility models (Hull and White, 1987; Heston and Nandi, 2000;

Nicolato and Venardos, 2003) as well as Markov switching (Campolieti and Makarov, 2005) and mixture models (Schittenkopf et al., 1998). However, none of these models are appropriate for the problem of estimating distributions of implied prices. In particular, these models assume that predictive distributions are conditionally normal, typically use a fixed number of mixture components for the mixtures and/or move all observations at a given time point simultaneously across components. These features seriously restrict the shape of the density estimates generated by the model, ruling out multimodal and skewed distributions. In contrast, our model provides support over a large class of continuous distributions, which allows it to capture the peculiar features of implied-price distribution. Also infinite mixtures allow us to automatically deal with the number of components in the mixture as a nuisance parameter, eliminating the need to select the number of components. Instead, estimates are obtained by averaging over different number of components according to their posterior probabilities.

We would like to note that inferring the distribution of implied prices discussed in this paper is different from estimating the risk-neutral distribution (Ait-Sahalia, 1996; Ait-Sahalia and Duarte, 2003; Panigirtzoglou and Skiadopoulos, 2004; Soderlind and Svensson, 1997), as the latter refers to the price at expiration while the former refers to the current price. Indeed, to the best of our knowledge, there is little precedent in the literature for the implied price distributions we are advocating in this paper.

## 2. Implied-price distributions in option markets

A *European call option* is an instrument that gives the buyer the right, but not the obligation, to buy an underlying asset at a fixed price $X$ (called the strike price) at a future time $T$ (called the expiration). Similarly, a *European put option* gives the buyer the right to sell an underlying asset at a fixed price $X$ at the expiration time $T$. For risk management and hedging purposes, options are usually traded in pairs.

Let $C_t(X)$ and $P_t(X)$ be, respectively, the price of a call and put option with strike value $X$ at time $t < T$. A well known no-arbitrage condition (Hull, 2005) states that

$$(1) \qquad C_t(X) - P_t(X) = S_t - X \exp\left(-r_t(T - t)\right)$$

where $S_t$ is the current spot price of the underlying asset and $r_t$ is the interest rate available to market actors at time $t < T$. Given the prices $C_t(X), P_t(X)$ we can solve for $S_t$ to obtain the *option-implied price* for the underlying asset,

$$(2) \qquad S_t^i(X) = C_t(X) - P_t(X) + X \exp\left(-r_t(T - t)\right)$$

As an illustration, consider the implied prices of the S&P500 between January 4th 1993 and March 17th 1994 depicted in Figure 1. We concentrate on options with three-month maturity and use the LIBOR as the interest rate for all our calculations (Panigirtzoglou and Skiadopoulos, 2004). The data set was constructed by Yacine Ait-Sahalia and has been used in other empirical studies (Duffie et al., 2000). The subset we employ contains a total of $n = 4385$ trades spread over $T = 306$ days, with sample sizes in any specific day varying between 0 and 26. Note that the distribution of prices on any specific day may be highly skewed and may have very heavy tails. Simple means (corresponding to the continuous line) vary wildly, specially during the summer of 1993 when fewer trades occur and extreme values are highly influential.

Option-implied prices obtained from (2) may dramatically differ from the spot price prevalent at the time. Indeed, the option-implied prices reveal the value that the actors in the derivatives market assign to the underlying asset based on their private information, which might be different from the information available to other investors. Figure 2 shows the differences between implied and spot prices in our sample, $\Delta_t = S_t^i - S_t$. Just as before, the distributions of expectations are left skewed and possibly multimodal, but no trend is apparent in this case. Implied prices tend to be about 7 points lower than the prevalent market price, which represents about a 1.5% difference, something reasonable when taking into account the transaction costs involved to synthetically replicate the underlying by taking long or short positions in call and put options as well as a zero coupon bond. However, in some cases the estimated differences can be as large as 15% of the spot price.

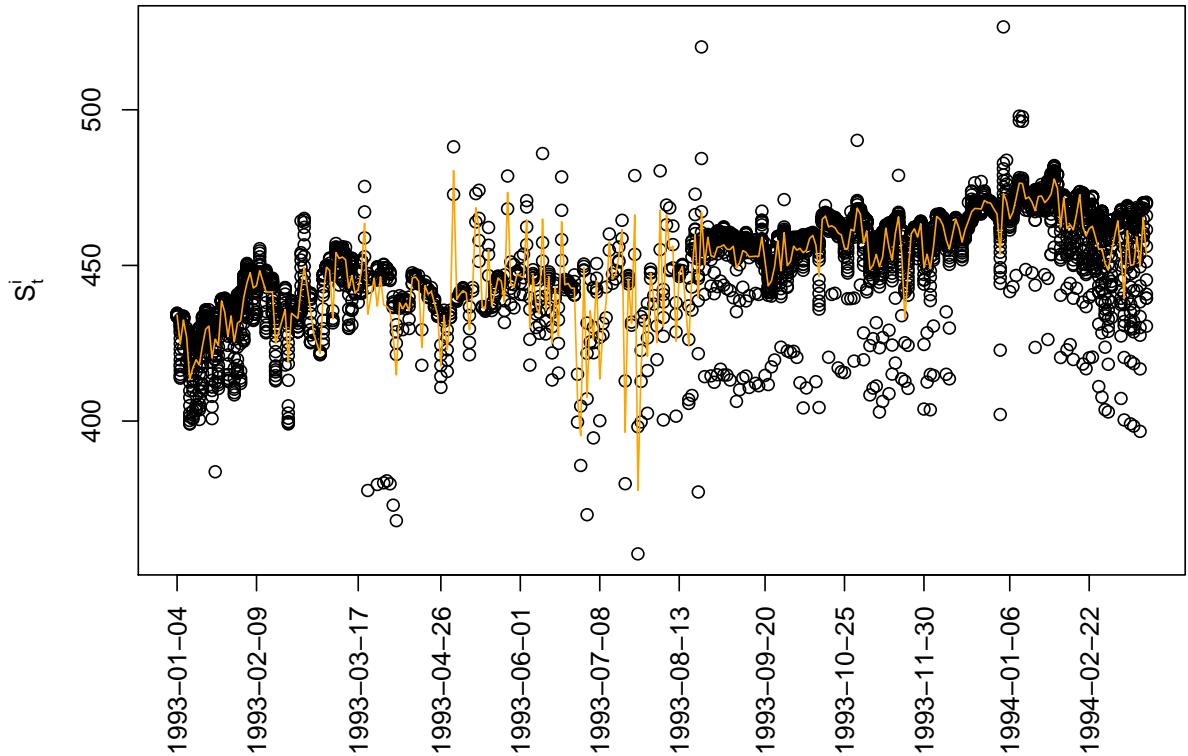

FIGURE 1. Implied prices in the S&P500 between January 4th 1993 and March 17th 1994. Multiple observations for any given day correspond to the different strike prices. Raw data is represented with dots, while the continuous line shows the evolution of daily averages in time.

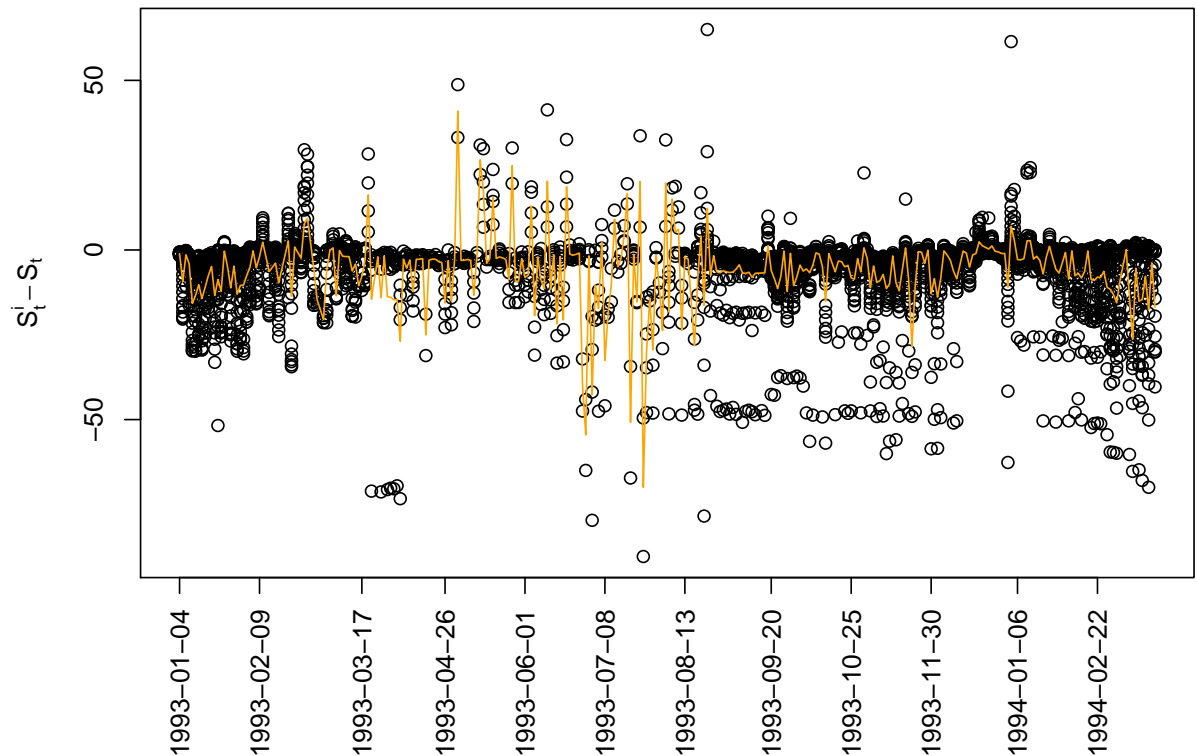

FIGURE 2. Differences between spot and implied prices between January 4th 1993 and March 17th 1994. The continuous line corresponds to daily means.

Figure 2 also helps us illustrate the fact that the distribution of implied prices is not the risk-neutral distribution, as it corresponds to the current price $S_t$ and not the prices at expiration, $S_T$. Indeed, note that the price of the option under the risk neutral density has to converge (in probability) to the market price as we approach expiration. This property is clearly not satisfied by the prices shown.

Some additional information on these distributions is provided Figure 3, which shows four consecutive kernel density estimates for the price differences $\Delta_t$ between February 18 and February 25, 1993 (no observations are available at February 19 and 22). Bandwidths were estimated independently at each time using crossvalidation (Silverman, 1986). Distributions can change dramatically but are generally multimodal and skewed. Bandwidth also change dramatically. From this, it is clear that standard parametric models are not a viable alternative in this setting, making non or semi-parametric methods a necessity.

## 3. BAYESIAN NONPARAMETRIC MODELING FOR COLLECTIONS OF TIME-EVOLVING DISTRIBUTIONS

In this Section, we discuss the statistical model for dynamic density estimation in the context of implied-price distributions. We start by reviewing the Dirichlet process and some of its extensions, and then move to discuss our model and its computational implementation.

3.1. **The Dirichlet process.** Let $(\mathcal{X}, \mathcal{B})$ be a complete and separable metric space (typically $\mathcal{X} = \mathbb{R}^n$ and $\mathcal{B}$ are the Borel sets on $\mathbb{R}^n$), and let $K \in \mathcal{K}$ be its associated probability measure. A Dirichlet process (Ferguson, 1973, 1974) with baseline measure $K_0$ and precision $\alpha$, denoted $\mathsf{DP}(\alpha K_0)$, defines a distribution on the space of probability measures $\mathcal{K}$, such that all distribution $K \sim \mathsf{DP}(\alpha K_0)$ if and only if it admits a representation of the form,

$$(3) \qquad K(\cdot) = \sum_{l=1}^{\infty} w_l^* \delta_{\boldsymbol{\eta}_l^*}(\cdot)$$

where $\{\boldsymbol{\eta}_l^*\}_{l=1}^{\infty}$ are independent and identically distributed samples from $K_0$ and $w_l^* = z_l^* \prod_{k=1}^{l-1}(1-z_k^*)$ with $\{z_l^*\}_{l=1}^{\infty}$ iid samples from a $\mathsf{Beta}(1, \alpha)$. Equation (3) is called the stick-breaking representation (Sethuraman, 1994), and it readily shows that the Dirichlet process places probability one on the subspace of discrete distributions.

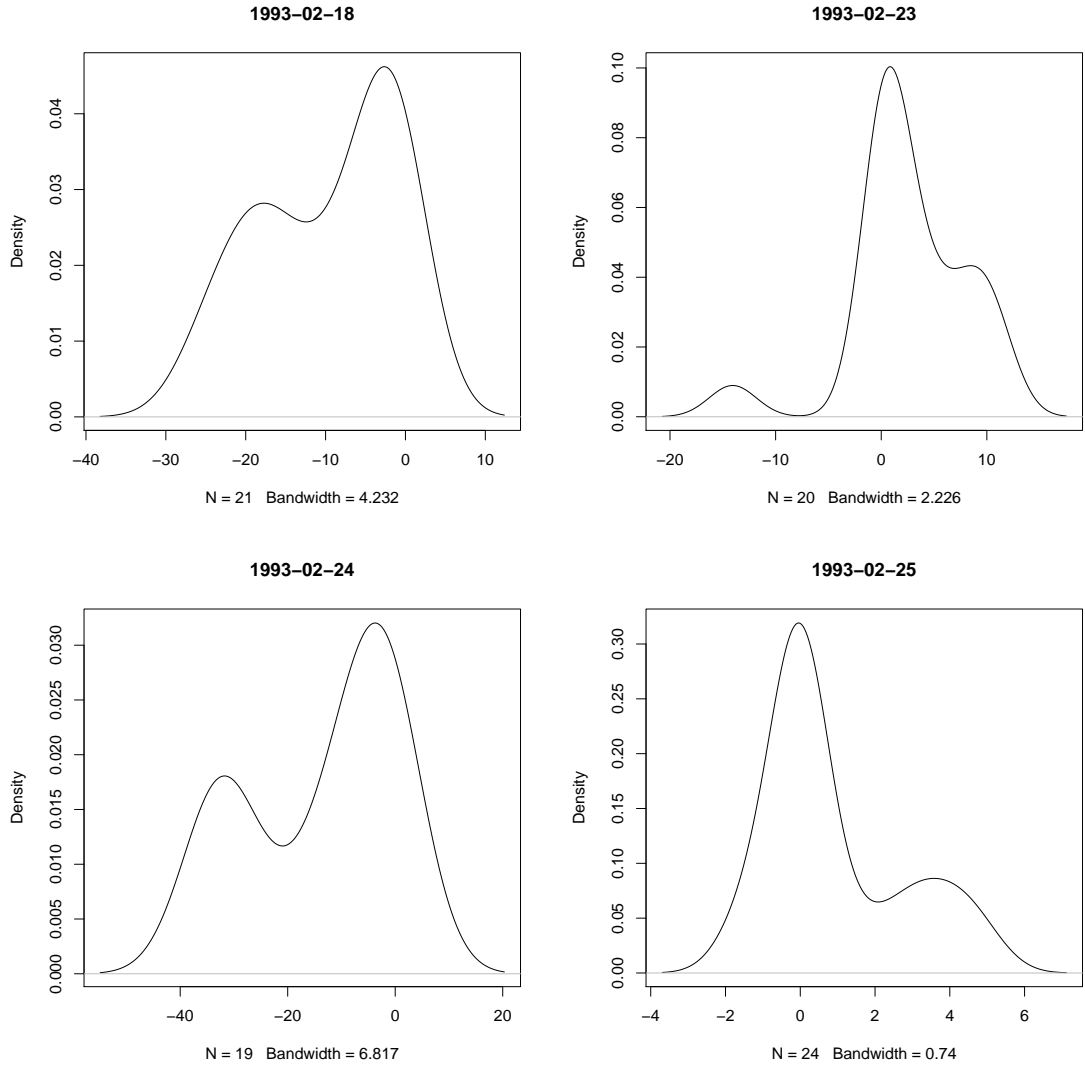

FIGURE 3. Kernel density estimates of implied differences in the S&P500 prices between February 19 and February 25, 1993. The number of observations $N$ and the bandwidth estimated through cross-validation are shown bellow each plot.

Another consequence of (3) is that, for any set $B \in \mathcal{B}$, $K(B)$ is a random quantity following a Beta distribution with parameters $\alpha K(B)$ and $\alpha(1 - K(B))$. In particular,

$$\mathbb{E}(K(B)) = K_0(B) \qquad \mathbb{V}(K(B)) = \frac{K_0(B)(1 - K_0(B))}{\alpha + 1}$$

Therefore, $K_0$ and $\alpha$ can be interpreted, respectively, as mean and precision parameters. In order to make the model more flexible, the DP is typically used to describe the (unknown) distribution of the parameters of some continuous distribution $H(\cdot|\boldsymbol{\eta})$, leading to the well known Dirichlet process mixture (DPM) models (Lo, 1984; Escobar, 1994):

$$y \sim \int H(y|\boldsymbol{\eta})K(d\boldsymbol{\eta}) \qquad\qquad K \sim \mathsf{DP}(\alpha K_0)$$

A common choice is $H(\cdot|\boldsymbol{\eta}) = \mathsf{N}(\cdot|\boldsymbol{\eta} = (\mu, \sigma^2))$, yielding a Gaussian location-scale mixture model that is dense in the space of absolutely continuous distributions (Lo, 1984). This model can also be interpreted as a Bayesian kernel density estimator, while the sequence $\{\mu_l^*\}_{l=1}^{\infty}$ controls the location of the Gaussian kernels, the set $\{\sigma_l^{*2}\}_{l=1}^{\infty}$ controls the bandwidth associated with each of them (Escobar, 1994).

The Dirichlet process is the most widely used nonparametric model for random distributions in Bayesian statistics; some recent applications include finance (Kacperczyk et al., 2003), econometrics (Chib and Hamilton, 2002; Hirano, 2002), epidemiology (Dunson, 2005), genetics (Medvedovic and Sivaganesan, 2002; Dunson et al., 2005), medicine (Kottas et al., 2002; Bigelow and Dunson, 2005) and auditing (Laws and O'Hagan, 2002). One of the main reasons for its popularity is the availability of efficient computational techniques (Neal (2000) provides an excellent review). Simulation algorithms for the Dirichlet process can be broadly divided in three groups: marginal samplers, which integrate out the unknown distribution $K$ (Escobar and West, 1995; MacEachern and Müller, 1998); blocked samplers, which exploit the stick-breaking construction in (3) (Ishwaran and James, 2001; Ishwaran and Zarepour, 2002; Ishwaran and James, 2002); and Reversible Jump samplers (Jain and Neal, 2000). In particular we focus our attention on marginal samplers, which exploit the Pòlya Urn representation of the predictive distribution of the process (Blackwell and MacQueen, 1973),

$$(4) \quad \boldsymbol{\eta}_n | \boldsymbol{\eta}_{n-1}, \ldots, \boldsymbol{\eta}_1 \sim \sum_{l<n} \frac{1}{\alpha+n-1} \delta_{\boldsymbol{\eta}_l} + \frac{\alpha}{\alpha+n-1} K_0$$

Due to the exchangeability of the observations, equation (4) also describes the full conditional distributions of any parameter given the rest, which are the basic ingredient required to develop Markov chain Monte Carlo schemes to sample from the posterior distribution of this model (details will be provided in Section 4).

3.2. **Dependent Dirichlet processes in discrete time.** Dirichlet process mixture models are a natural option if we are interested in a nonparametric model for a single probability measure. However, in many problems we are interested in how a distribution varies with another variable $\mathbf{s} \in S$; for example, when estimating expectations from option markets we would like to know how the distribution of option-implied prices changes as time evolves, as well as borrowing information across consecutive periods. A lot of recent attention has focused on extending the DP to collections of distributions on an index space $S$. One possible strategy is to introduce dependence through linear combinations of realizations of independent Dirichlet processes; some examples include Müller et al. (2004), Dunson (2006), Griffin and Steel (2006) and Dunson et al. (2004). Another alternative is to replace the elements in the stick-breaking representation (3) with sample paths from appropriate stochastic process in $S$ such that

$$K_{\mathbf{s}}(\cdot) = \sum_{l=1}^{\infty} w_l(\mathbf{s}) \delta_{\boldsymbol{\eta}_l^*(\mathbf{s})}(\cdot)$$

where $w_l(\mathbf{s}) = z_l^*(\mathbf{s}) \prod_{k=1}^{l-1}(1 - z_k^*(\mathbf{s}))$, $\{z_l(\mathbf{s})\}_{l=1}^{\infty}$ are independent and identically distributed sample paths from a stochastic process $\{z_l(\mathbf{s}) : \mathbf{s} \in S\}$ such that $z_l(\mathbf{s}) \sim \text{Beta}(1, \alpha(\mathbf{s}))$ for all $\mathbf{s}$, and $\{\boldsymbol{\eta}_l^*(\mathbf{s})\}_{l=1}^{\infty}$ are independent and identically distributed sample paths from another stochastic process $\{\boldsymbol{\eta}(\mathbf{s}) : \mathbf{s} \in S\}$. Note that, for any fixed $\mathbf{s}$, the measure $K_{\mathbf{s}}$ follows a regular DP. A construction of this type is called a dependent Dirichlet process (DDP) (MacEachern, 2000). Constant weight models, where the set of weights $\{w_l(\mathbf{s})\}_{l=1}^{\infty}$ are independent of $\mathbf{s}$, are one particularly useful subclass of DDPs as they allow for straightforward computational implementation that exploits (4). These have been used by DeIorio et al. (2004) to derive ANOVA models for distributions, Gelfand et al. (2005) to creative spatially varying nonparametric priors and by Rodriguez and ter Horst (2008) to construct dynamic density estimation models.

In the sequel, let $\Delta_{it}$ be the $i$-th difference between the implied price and the corresponding spot price observed at day $t$, for $t = 1, \ldots, T$ and $i = 1, \ldots, n_t$. For any fixed $t$, consider location mixtures of Gaussian distributions,

$$
\Delta_{it} \sim H = \int \mathsf{N}\left(\Delta_{it} | \mathbf{F}_{it}\boldsymbol{\theta}_t^*, \sigma_t^2\right) K_t(d\boldsymbol{\theta}_t)
$$
(5)
$$
K_t(\cdot) = \sum_{l=1}^{\infty} w_l^* \delta_{\boldsymbol{\theta}_{lt}^*}(\cdot)
$$

where $w_l^* = z_l^* \prod_{k=1}^{l-1}(1 - z_k^*)$, $z_l^* \sim \mathsf{Beta}(1, \alpha)$ and $\mathbf{F}_{it}$ is a given row vector. For a fixed time $t$, this mixture can be interpreted as a Gaussian kernel density estimate with common bandwidth $\sigma_t^2$ for all kernels. The relative importance of the kernels is controlled by the mass parameter $\alpha$; letting $\alpha \to 0$ implies that a single kernel is used and therefore we revert to a parametric (Gaussian) model. Hence, we see the mixture as a mechanism to approximate an unknown distribution.

Setting $\{\boldsymbol{\theta}_{lt}^*\}_{l=1}^{\infty}$ and $\sigma_t^2$ independent from $\{\boldsymbol{\theta}_{lt'}^*\}_{l=1}^{\infty}$ and $\sigma_{t'}^2$ would lead to density estimates that are independent a posteriori. Since we are interested in borrowing information across time, we introduce dependence by letting the location of the kernels evolve according to

(6) $$\boldsymbol{\theta}_{lt}^* | \boldsymbol{\theta}_{l,t-1}^* \sim \mathsf{N}(\mathbf{G}_t \boldsymbol{\theta}_{l,t-1}^*, \mathbf{W}_t) \qquad\qquad \boldsymbol{\theta}_{0l}^* \sim \mathsf{N}(\mathbf{m}_0, \mathbf{C}_0).$$

We also allow the bandwidth to adapt in time by setting

$$\sigma_t^2 = \frac{\zeta_t \sigma_{t-1}^2}{\delta} \qquad\qquad \zeta_t \sim \mathsf{Beta}(\frac{\delta n_t}{2}, \frac{(1-\delta)n_t}{2})$$

with initial condition $\sigma_0^2 \sim \mathsf{IG}(s_0, s_0 S_0)$. As in standard dynamic linear models (Carter and Kohn, 1994; West and Harrison, 1997), the evolution of the atoms follows a multivariate random walk. By appropriately choosing $\mathbf{F}_{it}$, $\mathbf{G}_t$ and $\mathbf{W}_t$, the model can accommodate trends, periodicities, autoregressions and dynamic regression models for the location of the kernels. In particular, $\mathbf{W}_t$ controls the magnitude of the change; letting $\mathbf{W}_t \to \mathbf{0}$ implies that the location of the kernels is the same at every time point. On the other hand, the stochastic volatility component follows the first order autoregressive process developed in Uhlig (1997). The discount factor $\delta$ controls the size of the change between adjacent days. Setting $\delta = 1$ yields a model with constant bandwidth, while lower values of $\delta$ yield models with less bandwidth smoothing. Since the level of smoothness in the process depends so critically on $\mathbf{W}_t$ and $\delta$, in the sequel we treat these parameters as unknown and estimate them along with all other parameters in the model.

This model can also be interpreted as an infinite mixture of linear filters with common evolution parameters but different state parameters (Rodriguez and ter Horst, 2008); as $\alpha \to 0$ the model reduces to a regular Kalman filter. Under this interpretation, we can rearrange all terms corresponding to the $l$-th mixture component to construct $\boldsymbol{\Theta}_l^* = (\boldsymbol{\theta}_{l0}^*, \boldsymbol{\theta}_{l1}^*, \ldots, \boldsymbol{\theta}_{lT}^*)'$, the $l$-th vector of state parameters. Although multiple observations from different time points can be assigned to the same mixture component, $y_{it}$ observation depends on the corresponding $\boldsymbol{\Theta}_l^*$ only through $\boldsymbol{\theta}_{lt}^*$. This representation will be exploited in Section 4 to develop a computational algorithm for the model.

The DDP model described above is rich enough to capture the features of the data highlighted in Section 2. In particular, it allows for multimodality, stochastic volatility and changing bandwidth (smoothness) in the density estimates. Also, since the no-arbitrage condition makes no assumption about the distribution of prices, our approach is more flexible than those based on implied volatilities.

In order to illustrate the flexibility of the model, consider the moments of the time varying distributions. Conditional on the mixing distribution $K_t$ we have,

$$\mathbb{E}(y_{it}|K_t) = \mathbf{F}'_{it}\left[\sum_{l=1}^{\infty} w_l^* \boldsymbol{\theta}_{lt}^*\right]$$

$$\mathbb{V}(y_{it}|K_t) = \sigma_t^{*2} + \mathbf{F}'_{it}\left[\sum_{l=1}^{\infty} w_l^* \boldsymbol{\theta}_{lt}^* \boldsymbol{\theta}_{lt}^{*'} - \left\{\sum_{l=1}^{\infty} w_l^* \boldsymbol{\theta}_{lt}^*\right\}\left\{\sum_{l=1}^{\infty} w_l^* \boldsymbol{\theta}_{lt}^{*'}\right\}\right]\mathbf{F}_{it}$$

$$\mathbb{C}\text{ov}(y_{it}, y_{i',t+k}|K_t) = \mathbf{F}'_{it}\left[\sum_{l=1}^{\infty} w_l^* \boldsymbol{\theta}_{lt}^* \boldsymbol{\theta}_{l,t+k}^{*'} - \left\{\sum_{l=1}^{\infty} w_l^* \boldsymbol{\theta}_{lt}^*\right\}\left\{\sum_{l=1}^{\infty} w_l^* \boldsymbol{\theta}_{l,t+k}^{*'}\right\}\right]\mathbf{F}_{i',t+k} \quad i \neq i'$$

These expressions show that the process is in general nonstationary; in particular, both the mean and the variance of the estimated distributions evolve in time. It is also possible to integrate out the unknown distribution $K_t$ under the Dirichlet process prior, which yields

$$\mathbb{E}(y_{it}) = \mathbf{F}'_{it}\mathbb{E}(\boldsymbol{\theta}_t)$$

$$\mathbb{V}(y_{it}) = \frac{1}{1+\alpha}\mathbf{F}'_{it}\mathbb{V}(\boldsymbol{\theta}_t)\mathbf{F}_{it} + \mathbb{E}(\sigma_t^2)$$

$$\mathbb{C}\text{ov}(y_{it}, y_{i',t+k}) = \frac{1}{1+\alpha}\mathbf{F}'_{it}\left[\prod_{s=1}^{k} \mathbf{G}_{t+k-s+1}\right]\mathbb{V}(\boldsymbol{\theta}_t)\mathbf{F}_{i',t+k} \quad i \neq i'$$

where $\mathbb{E}(\boldsymbol{\theta}_t)$, $\mathbb{V}(\boldsymbol{\theta}_t)$ and $\mathbb{E}(\sigma_t^2)$ can be obtained from the moments of the baseline measure,

$$\mathbb{E}(\boldsymbol{\theta}_t) = \left[\prod_{r=1}^{t} \mathbf{G}_{t-r+1}\right] \mathbf{m}_0$$

$$\mathbb{V}(\boldsymbol{\theta}_t) = \left[\prod_{r=1}^{t} \mathbf{G}_{t-r+1}\right] \mathbf{C}_0 \left[\prod_{r=1}^{t} \mathbf{G}_{t-r+1}\right]' +$$

$$\sum_{r=1}^{t-1} \left[\prod_{s=1}^{t-r} \mathbf{G}_{t-s+1}\right] \mathbf{W}_r \left[\prod_{s=1}^{t-r} \mathbf{G}_{t-s+1}\right]' + \mathbf{W}_t$$

$$\mathbb{E}(\sigma_t^2) = \left[\prod_{r=1}^{t} \frac{\delta n_r - \delta}{\delta n_r - 1}\right] \frac{s_0}{(s_0 - 1)} S_0 \quad n_t > 1$$

Therefore, if the evolution process for the atoms is stationary and $\mathbf{F}_{it}$ is constant for every $i$ and $t$, the resulting model for the distributions is a priori centered around a stationary process (even if it nonstationary a posteriori). The model described in this section is similar in spirit to that developed in Rodriguez and ter Horst (2008); the main difference is in the treatment of the variance of the Gaussian components. While Rodriguez and ter Horst (2008) assume that each component has a different variance, its value is fixed in time. This yields models that allow for the variance of the distribution to evolve in time in very restrictive ways. In contrast, the model described above keeps the variance constant across components but allows it to vary in time, endowing the model with greater flexibility.

For the application discussed in Section 5 we further specialize the model by setting $\mathbf{F}_{it} = 1$ and using a stationary first order autoregressive process for the evolution of the kernel locations,

$$K_t = \sum_{l=1}^{\infty} w_l^* \delta_{\theta_{lt}^*} \qquad \theta_{lt}^* | \mu, \theta_{l,t-1}^* \sim \mathsf{N}(\mu + \rho(\theta_{l,t-1}^* - \mu), U) \qquad \theta_{0l}^* \sim \mathsf{N}\left(\mu, \frac{U}{1 - \rho}\right)$$

This is a stochastic volatlity, distributional autoregressive model. The model is completed by establishing priors for the hyperparameters in the model. We give $\rho$ a $\mathsf{N}(0, 1)$ distribution truncated to the interval $(-1, 1)$ in order to ensure that the model is centered around a stationarity process. The evolution variance $U$ is assigned a conditionally conjugate inverse-gamma prior, $U \sim \mathsf{IG}(a_U/2, b_U/2)$. The

16    ABEL RODRÍGUEZ AND ENRIQUE TER HORST

discount factor $\delta$ is assigned a uniform prior on the the discrete set $\{0.4, 0.5, 0.6, 0.7, 0.8, 0.9, 0.95, 0.95\}$ for computational simplicity. Finally, the global mean $\mu$ is given a $\mathsf{N}(\mu_0, \tau^2)$ prior.

## 4. COMPUTATION

The use of simulation algorithms to fit Bayesian models has become commonplace in the last 15 years. In particular, Markov chain Monte Carlo (MCMC) algorithms, which generate a sequence of dependent samples from the posterior distribution of the parameters conditional on the data, are specially widespread. Given a starting guess for the value of the model parameters, these algorithm proceed iteratively by sampling from the full conditional distribtuion of blocks of parameters given all others in the model.

In this section we present a MCMC sampling scheme that exploits the Pòlya urn representation in (4) and the reformulation of the model as a mixture of linear filters. In the sequel, let $L$ be the current number of components in the mixture (5) that have observations allocated to them, $n_{lt}^*$ be the number of observations at time $t$ assigned to group $l$, $n_l^* = \sum_t n_{lt}^*$, $\boldsymbol{\Theta} = \{\boldsymbol{\Theta}_1^*, ..., \boldsymbol{\Theta}_L^*\}$, and $\boldsymbol{\Sigma} = \{\sigma_0^2, ..., \sigma_T^2\}$ be the current estimated values for those paths and time-varying variances. Also, $\xi_{it} = l$ iff $\boldsymbol{\theta}_{it} = \boldsymbol{\theta}_{lt}^\star$ and take negative superscripts to represent the corresponding vector *excluding* the relevant variables. Given values for the structural parameters $\mathbf{F}_{it}$, $\mathbf{G}_{it}$ and $\mathbf{W}_{it}$ and after initialization of the parameters, an MCMC sampler alternates through the following steps:

(1) For every $l = 1, \ldots, L$, generate $\boldsymbol{\Theta}_l^\star | \cdots$ using the following FFBS algorithm
   (a) Forward filter using the following recursions

$$\mathbf{m}_{lt} = \begin{cases} \mathbf{a}_{lt} + \mathbf{A}_{lt} e_{lt} & \text{if } n_{lt}^* > 0 \\ \mathbf{a}_{lt} & \text{if } n_{lt}^* = 0 \end{cases}$$

$$\mathbf{C}_{lt} = \begin{cases} \mathbf{R}_t - \mathbf{A}_{lt}\mathbf{Q}_{lt}\mathbf{A}'_{lt} & \text{if } n^*_{lt} > 0 \\ \mathbf{R}_{lt} & \text{if } n^*_{lt} = 0 \end{cases}$$

$$\mathbf{A}_{lt} = \mathbf{R}_{lt}\mathbf{F}_{lt}\mathbf{Q}_{lt}^{-1}$$

$$\mathbf{e}_{lt} = \mathbf{y}_{lt} - \mathbf{f}_{lt}$$

$$\mathbf{f}_{lt} = \mathbf{F}'_{lt}\mathbf{a}_{lt}$$

$$\mathbf{Q}_{lt} = \mathbf{F}'_{lt}\mathbf{R}_{lt}\mathbf{F}_{lt} + \sigma_t^2\mathbf{I}$$

$$\mathbf{a}_{lt} = \mathbf{G}_{lt}\mathbf{m}_{l,t-1}$$

$$\mathbf{R}_{lt} = \mathbf{G}_{lt}\mathbf{C}_{l,t-1}\mathbf{G}'_{lt} + \mathbf{W}_{lt}$$

where $\mathbf{y}_{lt}$ is made of all observations assigned to group $l$ at time $t$ and $\mathbf{F}_{lt}$ is a matrix whose rows are the corresponding $\mathbf{F}_{it}$ vectors.

(b) Sample $\boldsymbol{\theta}_{lT}|\cdots$ from $N(\mathbf{m}_{lT}, \mathbf{C}_{lT})$. Then recursively sample $\boldsymbol{\theta}_{lt}|\boldsymbol{\theta}_{l,t+1}, \cdots$ from $N(\mathbf{d}_{lt}, \mathbf{D}_{lt})$ where

$$\mathbf{d}_{lt} = \mathbf{B}_{lt}(\boldsymbol{\theta}_{l,t+1} - \mathbf{a}_{l,t+1})$$

$$\mathbf{D}_{lt} = \mathbf{C}_{lt} - \mathbf{B}_{lt}\mathbf{R}_{l,t+1}\mathbf{B}'_{lt}$$

$$\mathbf{B}_{lt} = \mathbf{C}_{lt}\mathbf{G}_{t+1}\mathbf{R}_{l,t+1}^{-1}$$

(2) Generate the sequence of variances $\Sigma|\Theta, \cdots$ using another FFBS algorithm

(a) Forward filtering using the following recursions

$$s_t = \delta s_{t-1} + n_t$$

$$S_t = \frac{\delta S_{t-1} + \sum_{i=1}^{n_t}(\Delta_{it} - \mathbf{F}_{it}\boldsymbol{\theta}^\star_{\xi_{it},t})^2}{\delta s_{t-1} + n_t}$$

(b) Backward sample, starting with $\sigma_T^2 \sim \left(\frac{s_T}{2}, \frac{s_T S_T}{2}\right)$ and then letting

$$\sigma_{t-1}^2 | \sigma_t^2 = \frac{1}{\eta_{t-1} + \frac{\delta}{\sigma_t^2}}$$

for all $0 \leq t < T$ where

$$\eta_{t-1} \sim \mathsf{G}\left(\frac{(1-\delta)s_{t-1}}{2}, \frac{S_{t-1}}{2}\right)$$

(3) Sample $\xi_{ij} | \xi_{ij}^-, \cdots$ from a multinomial distribution with probabilities:

$$q_l = n_l^{*-} p(\Delta_{it} | \mathbf{y}^-, \xi_{ij}^- \ l = 1, \ldots, L^-)$$

$$= n_l^{*-} \mathbf{N}\left(\Delta_{it} | \mathbf{h}_{lt}, \mathbf{H}_{lt}\right)$$

$$q_{L+1} = \alpha p(\Delta_{it} | S_0)$$

$$= \alpha \mathbf{N}\left(\Delta_{it} | \mathbf{h}_{t0}, \mathbf{H}_{t0}\right)$$

As before, $\mathbf{h}_{lT} = \mathbf{m}_{lT}$, $\mathbf{H}_{lT} = \mathbf{C}_{lT}$ and

$$\mathbf{h}_{lt} = \mathbf{B}_{lt}\left(\mathbf{h}_{l,t+1} - \mathbf{a}_{t+1}\right)$$

$$\mathbf{H}_{lt} = \mathbf{C}_{lt} - \mathbf{B}_{lt}(\mathbf{H}_{l,t+1} - \mathbf{R}_{l,t+1})\mathbf{B}'_{lt}$$

4.1. **Dealing with unknown hyperparameters.** In our application, the value of the evolution hyperparameters $\mu$, $\rho$ and $U$, as well as the discount factor $\delta$, are not known a priori and need to be estimated from the data. MCMC methods in hierarchical Bayes models can easily deal with this type of problems, where the parameters defining the statistical model for the observables of interest are unknown; all that is needed is again the full conditional distribution of the unknown hyperparameters given all other parameters.

MEASURING EXPECTATIONS IN OPTIONS MARKETS: AN APPLICATION TO THE S&P500 INDEX    19In the specific setting of the DPM models, sampling is simplified by noting that the realizations $\Theta_1^*, \ldots, \Theta_L^*$ are independent and identically distributed samples from the baseline measure defined by the evolution equations in (6). For example, the full conditional distributions for $\mu$ and $U$ are

$$\mu|\cdots \sim \mathsf{N}\left(\left[LT + \frac{U}{\tau^2}\right]^{-1}\left[\sum_{l=1}^{L}\sum_{t=1}^{T}\left(\frac{\theta_{lt}^* - \rho\theta_{l,t-1}^*}{1-\rho}\right)\right]; \left[\frac{LT}{U} + \frac{1}{\tau^2}\right]^{-1}\right)$$

$$U|\cdots \sim \mathsf{IG}\left(\frac{a_U + LT}{2}; \frac{b_U + \sum_{l=1}^{L}\sum_{t=1}^{T}\left[(\theta_{lt}^* - \mu) - \rho(\theta_{l,t-1}^* - \mu)\right]^2}{2}\right)$$

A similar argument can be used to construct a sampler for the correlation parameter $\rho$ and the discount factor $\delta$. In this case, a conditionally conjugate distribution is not available and we use a fine discrete grid to simplify computation while maintaining flexibility.

4.2. **Smoothing and predicting density estimates.** The original goal of our analysis is to obtain density estimates that borrow information across different periods and predict the shape of the density in the future. Given $D_T$, which stands for all the information up to time $T$, and the variance $\sigma_t^2$, the optimal estimator for the density at time $t < T$ under squared error loss corresponds to the posterior predictive distribution,

$$(7) \quad \hat{h}_t(\cdot|\sigma_t^2, D_T) = \mathbb{E}\left[\int \mathsf{N}(\cdot|\mathbf{F}_t'\boldsymbol{\theta}_t, \sigma_t^2)K_t(d\boldsymbol{\theta}_t)\Big|D_T\right] = \int \mathsf{N}(\cdot|\mathbf{F}_t'\boldsymbol{\theta}_t, \sigma_t^2)\mathbb{E}\left[K_t(d\boldsymbol{\theta}_t)\big|D_T\right].$$

We call this a filtered density estimate; it represents the best estimate available using *all* information in the sample, both past and future. In the specific case of the nonparametric DLM models discussed above, equation (7) can be used to derive the density estimates

$$(8) \quad \hat{h}_t(y|D_T) = \int \left[\sum_{l=1}^{L}\frac{1}{\alpha+L}\mathsf{N}(y|\mathbf{F}_t'\boldsymbol{\theta}_{lt}^*, \sigma_t^2) + \frac{\alpha}{\alpha+L}\mathsf{N}\left(y|\mathbf{F}_t'\mathbf{h}_{0t}, \sigma_t^2 + \mathbf{F}_t'\mathbf{H}_{0t}\mathbf{F}_t\right)\right]$$
$$p(\Theta_1^*, \ldots, \Theta_L^*, \sigma_t^2|D_T)d\Theta_1^*\ldots d\Theta_L d\sigma_t^2$$

Given a sample from the posterior distribution of the parameters in the model (say, of size $R$), the integral in (8) can be easily evaluated for any value of $y$ using Monte-Carlo integration as

$$\hat{h}_t(y|D_T) \approx \sum_{r=1}^{R} \left[ \sum_{l=1}^{L} \frac{1}{\alpha + L} \mathsf{N}(y|\mathbf{F}_t'\boldsymbol{\theta}_{lt}^{*(r)}, (\sigma_t^{(r)})^2) + \frac{\alpha}{\alpha + L} \mathsf{N}\left(y|\mathbf{F}_t'\mathbf{h}_{0t}, (\sigma_t^{(r)})^2 + \mathbf{F}_t'\mathbf{H}_{0t}\mathbf{F}_t\right) \right] \quad (9)$$

where the $r$ superscript denotes the $r$-th sample for the corresponding parameter, $r = 1, \ldots, R$. The $k$-step ahead density predictions $\hat{h}_{t+k}(\cdot|D_t)$, corresponding to the best density estimate obtained *only* from past information, can be obtained in a similar way.

## 5. Illustration: Implied market expectation prices for the S&P500 index

In this Section we apply the model from Section 3 to the data introduced in Section 2. We assume $m_0 \sim \mathsf{N}(\eta, \kappa^2)$ where $\eta = -10.0$ and $\kappa^2 = 100.0$. This choice reflects approximately the location and dispersion of the data. However, results were similar under our sensitivity analysis, which included values of $\eta$ between -30.0 and 10.0 and values of $\kappa^2$ between 25 and 400. Prior parameters for $\sigma^2$ were chosen as $s_0 = 1.0$ and $S_0 = 10.0$, while $U$ and $\alpha$ were assigned priors $\mathsf{IG}(2.0, 1.0)$ and $\mathsf{G}(1.0, 1.0)$ respectively. Finally, for the global mean we set $\mu \sim \mathsf{N}(\mu_0, \nu^2)$, with $\mu_0 = 0$ and $\nu^2 = 25$. Again, results were robust to moderate changes in these prior parameters.

A variant of the MCMC sampler described in Section 4 was used to fit this model. All results are based on 20,000 iterations obtained after a burn-in period of 5,000 samples. No convergence problems were evident from inspection of trace plots. Formal assessment of convergence was done using the Gelman-Rubin test (Gelman and Rubin, 1992), which compares the variability within and between multiple runs of the sampler with overdispersed starting values.

Figure 5 shows density estimates generated by the model for the trading days between February 18 and February 25 of 1993. These include the dates for the simple estimates in Figure 3, along with two additional dates (February 19 and 20) for which data was unavailable. The plots also show the original observations in order to demonstrate the plausibility of the estimates. First, we note that estimates are dramatically different. This is not really surprising; kernel density estimates are well known to be unreliable for very small sample sizes like ours, and standard methods do not borrow information across time, favoring enormous difference in estimates across consecutive time points. Second, we note that, in spite of the differences and in agreement with the descriptive analysis in Section 2, density estimates show negative skewness, with a very heavy left tail. This indicates that a relatively large number of actors in the option market tend to seriously undervalue the assets. Unlike the volatility smile observed in studies of implied volatility, this conclusion is not an artifact of the statistical model but an actually feature of the behavior of market actors.

Figure 5 displays the sequences of means and medians for the densities estimated by the model. Note that 1) both tend to be slightly negative, oscillating between 0 and -10 index points (which is expected as there are transaction costs associated with the synthetic portfolio), and 2) the means tend to be slightly smaller than the corresponding medians (which indicates again that the distributions of differences between market and implied prices are left skewed). Another striking feature is the sharp contrast with Figure 2; our location estimates do not present the wild swings observed in simple averages or standard parametric (Gaussian) models. Therefore, the model indicates that, for the S&P500, the private valuation of the asset by the "average" investor roughly agrees with the valuation in the spot market. However, this does not mean that all investors agree with these spot prices. Figure 6 shows estimates of volatility for the distributions of differences; high levels of volatility are associated with disagreement across actors and vice versa. Since the distributions are skewed and possibly multimodal, we use the interquartile range rather than the variances to measure volatility. In this setup, the interquartile range can be interpreted as the difference between the valuations of the 25% most bullish and the 25% most bearish investors. The interquartile range tends to be small (below 5 S&P500 points), showing that most investors tend to agree most of the time. However, the plot reveals one period of very high volatility in the late summer of 1993, along with three periods of moderate volatility around January 1993, April 1993 and March 1994. This information is particularly interesting as large discrepancies between the beliefs of market actors are thought to influence traded volume and returns. Indeed a simple regression of these estimated volatilities against next-day trading volumes yields a relatively high correlation, around 0.65.

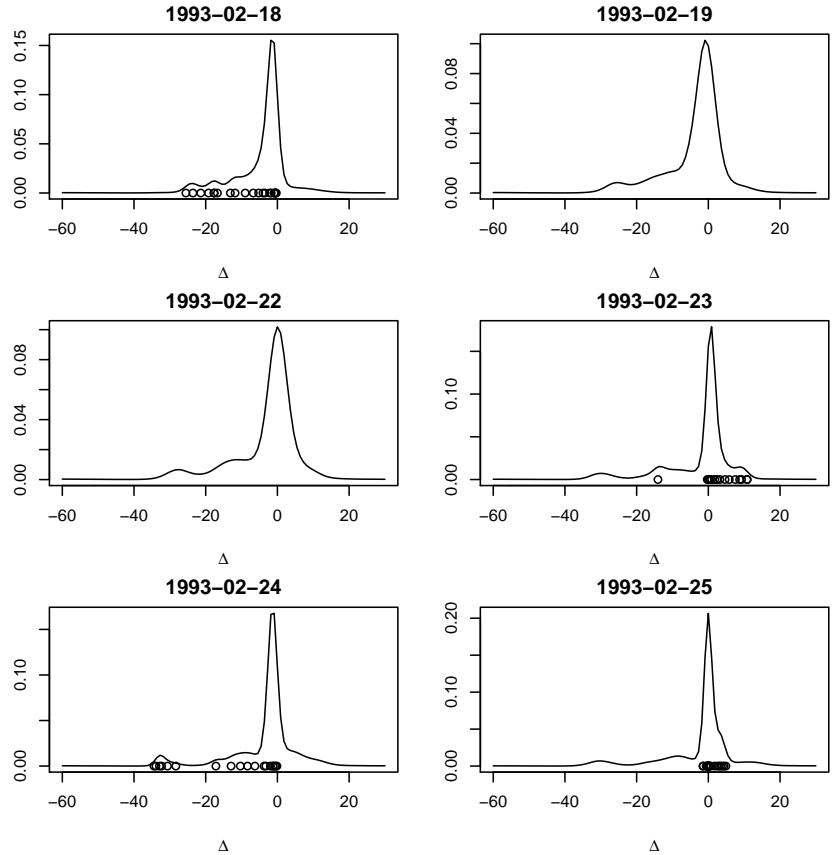

FIGURE 4. Density estimates for $\Delta_t$ (the difference between spot and implied prices) for all trading days between February 18 and February 25, 1993. Distributions show negative skewness and, in some cases, multimodality.

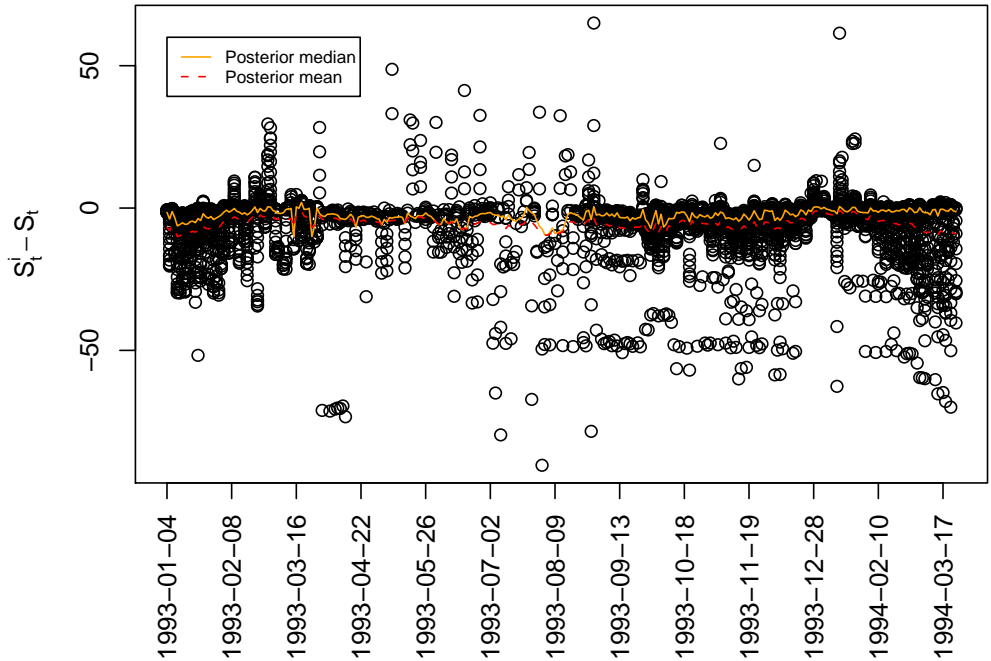

FIGURE 5. Estimates for the mean and median of the nonparametric density estimates. These can be interpreted as the expectations of the mean and median investor. Original observations are presented as reference.

Finally, another interesting measure of market expectations arising from our nonparametric density estimates is the proportion of market actors with extreme under/over valuation. As an illustration, we show in Figure 7 the probability that implicit valuations fall at least 15 points below the spot price. Since it is reasonable to assume that transaction costs for a very liquid market like the S&P500 are typically well below 3% of the price of the asset, this can be interpreted as the proportion of market actors that believe that the spot market is at least mildly overvalued. The results are quite interesting; although the typical proportion of investors that consider the stock market overvalued is between 5% and 10% most of the time, the proportion can jump to 25% in periods of serious disagreement. In contrast, a similar calculation on the upper tail (not shown) reveals that the proportion of market actors who think that the market is overvalued is consistently smaller even in periods high volatility.

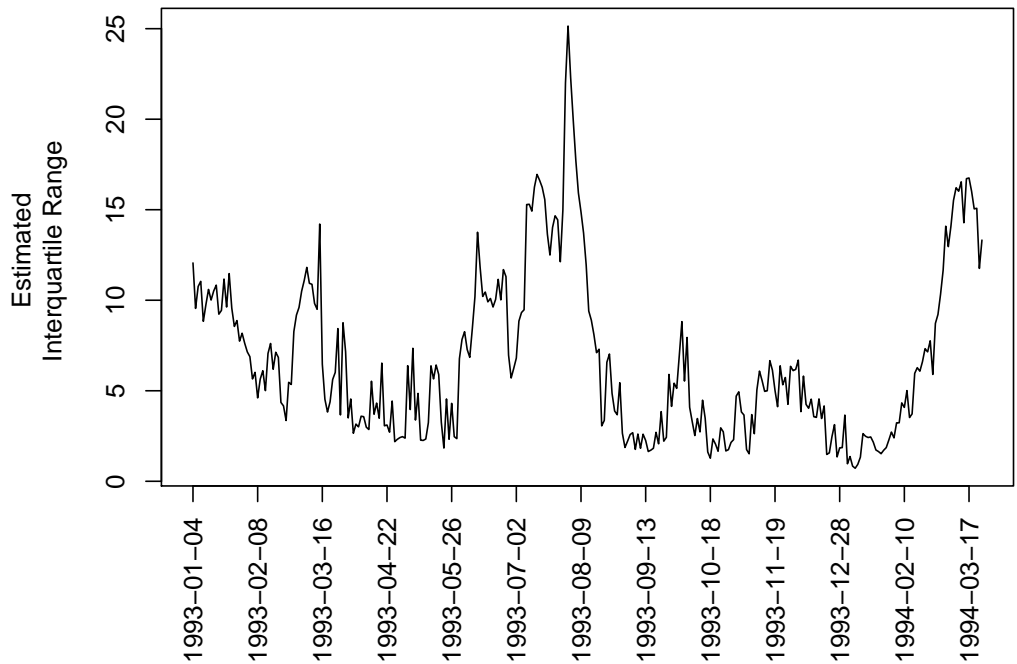

FIGURE 6. Interquartile range in the estimated densities. Large volatilities correspond to strong discrepancies across market actors with respect to the "fair" value of the asset.

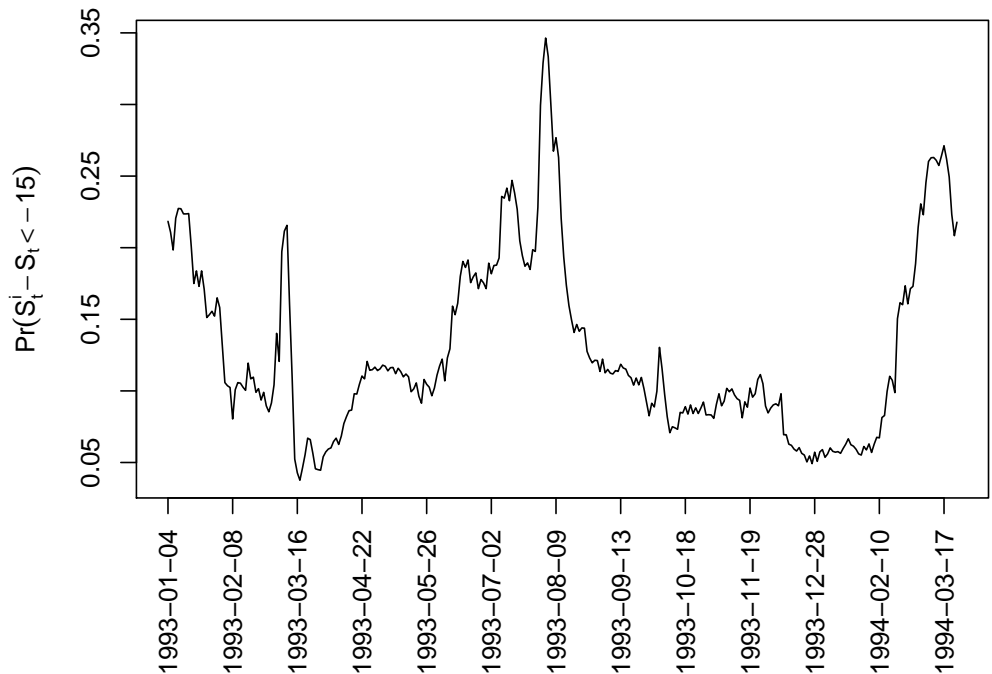

FIGURE 7. Probability that implicit valuations fall at least 15 points below the spot price. This can be interpreted as the proportion of market actors that believe that the spot market is overvalued.

6. DISCUSSION

We have demonstrated how information about asset prices contained in option markets can be recovered using nonparametric Bayesian methods. The model we propose is flexible enough to capture the characteristics of disributions of implied prices, which include skewness and multimodality, and does not rely on any of the parametric assumptions underlying the Black & Scholes valuation formula. The methodology provides dynamic estimates of the densities and, from them, we are able to derive different measures of expectations, including the behavior of the "mean" investor, the level of agreement across investors and the proportion of investors that think that the market is under/overpriced.

We illustrated our methodology using data from the S&P500 option market. In this case, the model indicates that the "average" investor in the option market has private valuations for the asset that are very close to the spot prices, with the consistently observed differences probably corresponding to transaction costs. This is not terribly surprising as both the spot and options markets for the S&P500 are extremely liquid. We expect that our methods can provide different results in more illiquid markets, where large differences between spot and implied prices can happen due to small sample sizes and information asymmetry.

The model also shows that, although most market actors tend to agree with the average investor on the price of the S&P500 index, valuations can substantially differ across investors, specially in periods of low market volumes. This points again to the utility of these methods in illiquid markets, where they can reveal mispricing issues. In addition, we have demonstrated that these periods of high uncertainty and disagreement among investors tend to be periods where a large proportion of actors consider that the index (and by extension, the stock market as a whole) is overvalued. Surprisingly, rarely over this period investors seem to think that the stock market is undervalued. This can be partially explained by the widespread practice of using the derivative market for hedging; bullish transactions in the spot market can be offset by bearish transactions in the option market. However, given the enormous size of derivative markets, we think this cannot be the full story and that underpricing indeed reveals private information.

Furthermore, the recent surge of a new methodological framework assessing sovereign credit risk (Gray et al., 2007; Gray and Malone, 2008) has opened new doors for policy makers not only to manage sovereign credit risk, but also to manage the risk of the sectors composing the economy of a country (households, financial and public). In this context, it is possible to extract from the balance sheet of the different sectors of an economy, the observations for the implied asset level (underlying) of the sector (Gray et al., 2007; Gray and Malone, 2008), over a given time period. This could allow us to study the evolution of the time-varying distribution of the asset level for a particular sector with the model developed in this manuscript, and thus to better understand the risk characteristics of the sector assets. This will be the theme of a future work.

Our description of the model ignores bid-ask spreads, whose presence yields *intervals* of rational implied prices rather than a single price. However, in our example, the spreads are so small that they can be safely overlooked and the interval replaced by its midpoint. In markets where spreads are noticeable, the model can be easily extended by imputing the true prices as part of the MCMC sampling scheme. Details will be provided elsewhere.

We already argued that our methods provides estimates for the distribution of current prices $S_t$, and not for the prices at expiration, $S_T$. Therefore the implied-price distributions we show in this paper do not correspond to implied risk-neutral distributions for the S&P500. However, if the process driving the implied prices is assumed to be a martingale (Delbaen and Schachermayer, 2006), then the expectation of the current price distribution is equal to the expectation of the prices at expiration and implied-price distributions can be used to provide information about the risk-neutral distribution.